\documentclass[11pt]{article}
\usepackage{amssymb}

\begin{document}

\title{ Supersymmetric Electroweak Corrections to the Chargino Decay into
       Neutralino and $W$ Boson
\footnote{Supported by National Natural Science Foundation of China.}}
\vspace{3mm}

\author{{ Zhang Ren-You$^{2}$,Ma Wen-Gan$^{1,2}$,Wan Lang-Hui$^{2}$ }\\
{\small $^{1}$ CCAST (World Laboratory), P.O.Box 8730, Beijing 100080, P.R.China} \\
{\small $^{2}$ Department of Modern Physics, University of Science and Technology}\\
{\small of China (USTC), Hefei, Anhui 230027, P.R.China} }

\date{}
\maketitle
\vskip 12mm

\begin{abstract}
We investigated the electroweak corrections from one-loop diagrams
involving the third generation (s)quarks to the decay width of
process $ \tilde{\chi}_1^+ \rightarrow W^+ \tilde{\chi}_1^{0} $ in
the framework of the Minimal Supersymmetric Standard Model. Our
calculation shows that these corrections are not very sensitive to
the mass of the lightest neutralino $\tilde{\chi}_1^{0}$ and the
sbottom mixing angle $\theta_{b}$, but depend strongly on the top
squark mixing angle $\theta_{t}$ and tan$\beta$. With our chosen
parameters, we find that these radiative corrections can exceed $
10 \%$, therefore they should be taken into account for the
precise experimental measurement at future colliders.
\end{abstract}

\vskip 5cm

{\large\bf PACS: 11.30.Pb, 12.15.Lk, 12.60.Jv, 14.80.Ly}

\vfill \eject

\baselineskip=0.36in

\renewcommand{\theequation}{\arabic{section}.\arabic{equation}}
\renewcommand{\thesection}{\Roman{section}}
\newcommand{\nb}{\nonumber}

\makeatletter      
\@addtoreset{equation}{section}
\makeatother       

\section{Introduction}
\par
Over the past years, many experimental works were focused on the
finding new physics beyond the standard model (SM)
\cite{s1}. At the same time, some new theoretical models have been
developed in order to describe the new physics, such as the
minimal supersymmetric model (MSSM) \cite{mssm1} \cite{mssm2}
\cite{mssm3}, large extra dimensions model \cite{add}, etc. Among
these models, the MSSM is arguably the most promising candidate
for extension beyond the SM,
in which all the SM particles have their supersymmetric partners.
The existence of these supersymmetric particles is one of the
characteristics of this theory. Therefore, searching for these
supersymmetric particles at colliders is an important task to
verify the MSSM.
In the MSSM there exist two charginos
$\tilde{\chi}_i^+$ (i = 1, 2) and four neutralinos
$\tilde{\chi}_i^0$ (i = 1, ..., 4). They are just the mixtures
of gauginos and higgsinos, the supersymmetric partners of gauge
bosons and Higgs bosons \cite{mssm1} \cite{mssm2}.
The lightest neutralinos $\tilde{\chi}^0_1, \tilde{\chi}^0_2$
and chargino $\tilde{\chi}^+_1$
($m_{\tilde{\chi}_1^+} < m_{\tilde{\chi}_2^+}$) are expected to
be the lightest supersymmetric particles. For R-parity conservation,
the lightest supersymmetric particle (LSP) $\tilde{\chi}^0_1$ is stable
and invisible. Thus, the states $\tilde{\chi}^0_2$ and $\tilde{\chi}^+_1$
might be the first supersymmetric particles to be discovered.
The search for $\tilde{\chi}^+_1$ and $\tilde{\chi}^0_2$ is the main
task at present and future colliders. Many experiments have been done
at present colliders, such as LEP, Tevatron, etc.,
in searching for the supersymmetric particles \cite{search sp-1}
\cite{search sp-2} \cite{search sp-3},
especially for charginos and neutralinos \cite{search ch-1}
\cite{search ch-2}.
At the same time, the phenomenological calculations of chargino and
neutralino productions via
$\gamma \gamma$, $e^- \gamma$, $e^+ e^-$, $e^- e^-$, $\gamma p$, $pp$
collisions, etc., have also been performed. They can be found in
Ref.\cite{rr1,rr2,e-r,e+e-,e-e-,rp,gg}.
\par
The decays of charginos and neutralinos in the MSSM have been
widely discussed \cite{decay-1} \cite{decay-2}. If the mass
splitting between $\tilde{\chi}^0_1$ (LSP) and $\tilde{\chi}^+_1$
or $\tilde{\chi}^0_2$ is not large enough, the three-body decay
modes of $\tilde{\chi}^+_1$ and $\tilde{\chi}^0_2$ into
$\tilde{\chi}^0_1$ and two fermions through vector boson, Higgs
boson and sfermion exchanges are the dominant decay modes, which
take place mainly in the minimal supergravity (MSUGRA) scenario.
But in the models without gaugino mass unification at the GUT
scale, the mass splitting between them can exceed the masses of
gauge bosons ($W, Z$) and Higgs bosons in some parameter space.
Therefore, the decay modes $\tilde{\chi}^+_1(\tilde{\chi}^0_2)
\rightarrow W^+(Z^0) + \tilde{\chi}^0_1$,
$\tilde{\chi}^+_1(\tilde{\chi}^0_2) \rightarrow H^+(h^0,H^0,A^0) +
\tilde{\chi}^0_1$ are kinematically accessible. In this case, the
above two-body decay channels will be the most important decay
modes of $\tilde{\chi}^+_1$ and $\tilde{\chi}^0_2$, since the
chargino and neutralino have enough phase space to decay firstly
via these channels. If $\tilde{\chi}^+_1$ and $\tilde{\chi}^0_2$
are rather heavy, some new decay channels, such as decay into $f
\tilde{f}$ pair, should be under consideration when they are
kinematically allowed \cite{decay-1}. In this paper, we will only
consider the decay of the lightest chargino into the lightest neutralino
and $W$ boson. As mentioned above, in the case of
$m_{\tilde{\chi}_1^+} > m_W + m_{\tilde{\chi}_1^0}$, this decay
channel can open up and will be the dominant decay channel of
$\tilde{\chi}^+_1$. The W boson's leptonic decay may be used as a
spectacular trigger. For the comparison of the theoretical
prediction with the precise measurement at the future colliders,
the more accurate calculation of the partial decay width of this
decay mode is necessary. For this purpose, the radiative
corrections at one-loop level should be included. Since there is
no one-loop QCD correction to this process, the electroweak
corrections from one-loop quark-squark diagrams are the most important. In
this paper, only the third generation quarks and squarks are
considered because of the Yukawa characteristic factor $m_q/m_W$
in the chargino/neutralino-quark-squark vertices ($m_q$ is the
mass of quark q)\cite{haber}. In our calculation we adopted the
t'Hooft gauge. The relevant diagrams of the tree-level, counter-term,
one-loop vertex corrections to this decay and the self-energies
of W-boson, charginos and neutralinos, are shown in
Fig.1(a), (b), (c), (d), (e) and (f) (i = 1, j = 1), respectively.
\par
The paper is organized as follows: In section 2, we review briefly
the chargino, neutralino and squark sectors of the MSSM and give
the tree-level results (amplitude and decay width). In section 3,
we give some analytical results of the electroweak corrections to
the $\tilde{\chi}_1^+ \rightarrow W^+ \tilde{\chi}_1^0$ partial
decay width. The numerical results and our conclusions are
presented in section 4.

\section{Notations and tree-level results}
The tree-level Lagrangian for the chargino and neutralino masses reads
\begin{equation}
\label{lagrangian of mass}
{\cal L}_m^{\tilde{\chi}} = - \frac{1}{2}
            \left(
                   \begin{array}{cc}
                   \psi^+ \\ \psi^-
                   \end{array}
            \right)^T
            \left(
                   \begin{array}{cc}
                   0 & X^T \\
                   X & 0
                   \end{array}
            \right)
            \left(
                  \begin{array}{cc}
                  \psi^+ \\
                  \psi^-
                  \end{array}
            \right)
             -\frac{1}{2}\psi^{0T} Y \psi^0
         +h.c.
  \end{equation}
where the tree-level mass matrices $X$, $Y$ are \cite{haber}
\begin{eqnarray}
\label{mxn}
X &=& \left(
          \begin{array}{cc}
                M & \sqrt{2}m_W \sin\beta\\
                \sqrt{2}m_W \cos\beta & \mu
          \end{array}
    \right) \nb \\
Y &=& \left(
          \begin{array}{cccc}
               M^{\prime} & 0 & -m_Z s_W \cos\beta & m_Z s_W \sin\beta  \\
               0 & M & m_Z c_W \cos\beta & -m_Z c_W \sin\beta \\
               -m_Z s_W \cos\beta & m_Z c_W \cos\beta & 0 & -\mu \\
               m_Z s_W \sin\beta & -m_Z c_W \sin\beta & -\mu & 0
          \end{array}
      \right)
\end{eqnarray}
and $\psi^{+T} = (-i \lambda^+, \psi_{H_2}^+)$, $\psi^{-T} = (-i
\lambda^-, \psi_{H_1}^-)$, $\psi^{0T} = (-i \lambda^{\prime}, -i
\lambda^3, \psi_{H_1}^0, \psi_{H_2}^0)$, where $s_W\equiv \sin
\theta_W, c_W \equiv \cos \theta_W$, $M$ and $M^{\prime}$ are the
$SU(2)$ and $U(1)_Y$ soft-SUSY-breaking parameters, respectively.
We define
\begin{eqnarray}
\chi^+ & = & V \psi^+ , \nb \\
\chi^- & = & U \psi^+ , \nb \\
\chi^0 & = & N \psi^0 ,
\end{eqnarray}
where $U$, $V$ and $N$ are unitary matrices chosen such that:
\begin{eqnarray}
U^*XV^{-1} &=& M_D ={\rm diag}(m_{\tilde{\chi}_1^+}, m_{\tilde{\chi}_2^+}),
~~~~~~~(0<m_{\tilde{\chi}_1^+}<m_{\tilde{\chi}_2^+}) \nb \\
N^*YN^{-1} &=& M^0_D = {\rm diag}(m_{\tilde{\chi}_1^0}, m_{\tilde{\chi}_2^0},
                                  m_{\tilde{\chi}_3^0}, m_{\tilde{\chi}_4^0}),
(0<m_{\tilde{\chi}_1^0}<m_{\tilde{\chi}_2^0}<m_{\tilde{\chi}_3^0}<
  m_{\tilde{\chi}_4^0}). \nb \\
\end{eqnarray}
The tree-level Lagrangian for the stop and sbottom masses is
\begin{equation}
{\cal L}_m^{\tilde{q}} =
         -\left(
         \begin{array}{c}
         \tilde{t}_L \\
         \tilde{t}_R
         \end{array}
          \right)^{\dag} {\cal M}_{\tilde{t}}^2
      \left(
         \begin{array}{c}
         \tilde{t}_L \\
         \tilde{t}_R
         \end{array}
          \right)
         -\left(
         \begin{array}{c}
         \tilde{b}_L \\
         \tilde{b}_R
         \end{array}
          \right)^{\dag} {\cal M}_{\tilde{b}}^2
      \left(
         \begin{array}{c}
         \tilde{b}_L \\
         \tilde{b}_R
         \end{array}
          \right)
\end{equation}
in which the tree-level stop and sbottom squared-mass matrices are:
\begin{eqnarray}
{\cal M}_{\tilde{t}}^2 &=&
          \left(
          \begin{array}{cc}
          M_{\tilde{Q}}^2+m_t^2+m_Z^2\cos 2\beta \left(\frac{1}{2}-
          \frac{2}{3}s_W^2\right) & m_t \left(A_t-\mu \cot \beta\right) \\
          m_t \left(A_t - \mu \cot \beta\right) & M_{\tilde{U}}^2+m_t^2+
          \frac{2}{3}m_Z^2\cos 2\beta s_W^2
          \end{array}
      \right) \nb \\
{\cal M}_{\tilde{b}}^2 &=& \left( \begin{array}{cc}  M_{\tilde{Q}}^2
+ m_b^2 - m_Z^2 \cos 2\beta \left(\frac{1}{2} - \frac{1}{3}
s_W^2\right) & m_b\left(A_b - \mu \tan \beta\right) \\ m_b
\left(A_b - \mu \tan \beta\right) &
M_{\tilde{D}}^2+m_b^2-\frac{1}{3}m_Z^2\cos 2\beta s_W^2
\end{array}\right)
\end{eqnarray}
where $M_{\tilde{Q}},M_{\tilde{U}}$ and $M_{\tilde{D}}$ are the
soft-SUSY-breaking masses. For the third generation SU(2) squark
doublet, we have $\tilde{Q}=\left(\tilde{t}_L,\tilde{b}_L\right)$,
and the singlets $\tilde{U}=\tilde{t}_R$ and
$\tilde{D}=\tilde{b}_R$, respectively. $A_{t,b}$ are the
corresponding soft-SUSY-breaking trilinear coupling parameters.
\par
To diagonalize the mass matrices ${\cal M}_{\tilde{t}}^2$ and
${\cal M}_{\tilde{b}}^2$, we should introduce two unitary matrices
${\cal R}^{\tilde{q}}$: ${\cal R}^{\tilde{q}} {\cal
M}_{\tilde{q}}^2 {\cal R}^{\tilde{q}+} = {\rm diag}(m_{q_1}^2,
m_{q_2}^2), (q = t, b)$. In consideration of the reality of the
mass matrices, we may choose matrices ${\cal R}^{\tilde{q}}$ to be
\begin{equation}
{\cal R}^{\tilde{q}} =
     \left(
          \begin{array}{cc}
          \cos \theta_{q} & \sin \theta_{q} \\
     -\sin \theta_{q} & \cos \theta_{q}
          \end{array}
     \right)
     ~~~~~~(q = t, b),
\end{equation}
where $-\pi/2 \le \theta_{\tilde{q}}\le \pi/2$. The
mass-eigenstates $\tilde{q}_1, \tilde{q}_2$ are related to the
current eigenstates $\tilde{q}_L, \tilde{q}_R$ by the
transformation:
\begin{equation}
\left(
    \begin{array}{c}
     \tilde{q}_1 \\
     \tilde{q}_2
    \end{array}
\right) = {\cal R}^{\tilde{q}}
\left(
    \begin{array}{c}
     \tilde{q}_L \\
     \tilde{q}_R
    \end{array}
\right).
\end{equation}
\par
The tree-level Lagrangian for the chargino-neutralino-W boson interactions
is \cite{mssm2}
\begin{equation}
\label{L for interaction}
{\cal L}_{\tilde{\chi}^+{\tilde{\chi}^0W}} =
   g \bar{\tilde{\chi}}_i^0 \gamma^{\mu}
      \left[O^L_{i j} P_L + O^R_{i j} P_R \right]
     \tilde{\chi}_j^+ W_{\mu}^- + h.c.
\end{equation}
where $P_{L,R} = \frac{1}{2}(1 \mp \gamma_5)$, and
\begin{eqnarray}
\label{definition of O}
O^L_{i j} &=& -\frac{1}{\sqrt{2}} N_{i 4} V_{j 2}^* + N_{i 2} V_{j 1}^* \nb \\
O^R_{i j} &=&  \frac{1}{\sqrt{2}} N_{i 3}^* U_{j 2} + N_{i 2}^* U_{j 1}
\end{eqnarray}
From this Lagrangian, we can easily obtain the tree-level amplitude and partial
width of the decay $\tilde{\chi}_1^+ \rightarrow W^+ \tilde{\chi}_1^0$:
\begin{eqnarray}
\label{width at tree level}
&& \Gamma_{tree}(\tilde{\chi}_1^+ \rightarrow W^+ \tilde{\chi}_1^0)
      = \frac{\lambda}{16 \pi m_{\tilde{\chi}_1^+}^3}
      \mid {\cal M}_{tree}\mid^2 \nb \\
 &&= \frac{\lambda g^2}{16 \pi m_{\tilde{\chi}_1^+}^3}
     \left[ \rho \left(\mid O^L_{1 1}\mid^2 + \mid O^R_{1 1}\mid^2\right)
    -12 m_{\tilde{\chi}_1^+} m_{\tilde{\chi}_1^0} {\rm Re}\left(O^{L *}_{1 1}
    O^R_{1 1}\right)
    \right]
\end{eqnarray}
where
\begin{eqnarray}
\lambda &=& \left[\left( m_{\tilde{\chi}_1^+}^2 -
            ( m_{\tilde{\chi}_1^0} + m_W )^2 \right)
        \left( m_{\tilde{\chi}_1^+}^2 -
            ( m_{\tilde{\chi}_1^0} - m_W )^2 \right)\right]^{1/2} \nb \\
\rho &=& m_{\tilde{\chi}_1^+}^2 + m_{\tilde{\chi}_1^0}^2 - 2 m_W^2
       + \frac{(m_{\tilde{\chi}_1^+}^2 - m_{\tilde{\chi}_1^0}^2)^2}{m_W^2}
\end{eqnarray}

\section{Renormalization and radiative corrections}
The supersymmetric radiative corrections to chargino and
neutralino productions have been considered widely \cite{kiyoura}
\cite{diaz}. For convenience, the corrections are expressed by
using form factors. All the one-loop level contributions to these
form factors can be classified in terms of prototypes
distinguished by the number of particles inside the loops and
their spin \cite{diaz}. The calculation of the supersymmetric
corrections to the decay $\tilde{\chi}^+_1 \rightarrow W^+
\tilde{\chi}^0_1$ considered in this paper can be performed
analogously. The form factors can be divided into two parts, which
are just the contributions of counter-term and vertex correction.
In order to obtain the counter-term of vertex
$\bar{\tilde{\chi}}_i^0 \tilde{\chi}_j^+ W^-$ which will be used
in the calculation of the corrections, the following
renormalization constants are defined \\
(1) renormalization constants of fields \cite{denner}:
\begin{eqnarray}
\label{field renor}
W_{\mu} & \rightarrow & Z_W^{1/2} W_{\mu} \nb \\
P_L \tilde{\chi}_i^+ & \rightarrow & Z^{L1/2}_{+,i j}P_L \tilde{\chi}_j^+,~~~~
P_R \tilde{\chi}_i^+  \rightarrow  Z^{R1/2}_{+,i j}P_R \tilde{\chi}_j^+ \nb \\
P_L \tilde{\chi}_i^0 & \rightarrow & Z^{L1/2}_{0,i j}P_L \tilde{\chi}_j^0,~~~~~
P_R \tilde{\chi}_i^0  \rightarrow  Z^{R1/2}_{0,i j}P_R \tilde{\chi}_j^0 \nb
\end{eqnarray}
(2) renormalization constants of gauge couplings and masses \cite{denner}:
\begin{eqnarray}
\label{para renor}
e & \rightarrow & Z_e e = (1 + \delta Z_e)e \nb \\
g & \rightarrow & g + \delta g = (1 + \delta Z_e - \frac{\delta s_W}{s_W})g \nb\\
m_W^2 & \rightarrow & m_W^2 + \delta m_W^2,~~~~m_Z^2 \rightarrow m_Z^2 \nb
+\delta m_Z^2
\end{eqnarray}
(3) renormalization of matrices $U, V$ and $N$ \cite{yamada2}:
\begin{equation}
\label{uvn renor} U \rightarrow U+\delta U,~~~V \rightarrow
V+\delta V,~~~ N \rightarrow N+\delta N,
\end{equation}
where $Z^R_{0,i j} = Z^{L*}_{0,i j}$, because neutralinos
$\tilde{\chi}_i^0$ are Majorana particles,
$\tilde{\chi}_i^{0c}=\tilde{\chi}_i^0$.
\par
In this paper, we adopt the complete on-mass-shell scheme in doing
renormalization. The relevant renormalization constants in eq.(\ref{uvn renor})
are fixed by the on-mass-shell renormalization conditions:
\par
(a) The on-mass-shell renormalization conditions for gauge bosons are \cite{denner}
\begin{eqnarray}
&& \lim_{k^2 \rightarrow m_W^2} \frac{1}{-i (k^2-m_W^2)} \tilde{Re}
  i \hat{\Gamma}_{\mu \nu}^{WW}(k)\epsilon^{\nu}(k)=\epsilon_{\mu}(k) \nb \\
&& \lim_{k^2 \rightarrow m_Z^2} \frac{1}{-i (k^2-m_Z^2)} \tilde{Re}
  i \hat{\Gamma}_{\mu \nu}^{ZZ}(k)\epsilon^{\nu}(k)=\epsilon_{\mu}(k) \nb \\
&& \lim_{k^2 \rightarrow 0} \frac{1}{-i k^2} \tilde{Re}
  i \hat{\Gamma}_{\mu \nu}^{\gamma \gamma}(k)\epsilon^{\nu}(k)=\epsilon_{\mu}(k) \nb \\
&& \tilde{Re}i \hat{\Gamma}_{\mu \nu}^{\gamma Z}(k)\epsilon^{\nu}(k)\mid_{k^2=0, m_Z^2}=0
\end{eqnarray}
where $\epsilon(k)$ are the polarization vectors of the external
fields. $\tilde{Re}$ takes only the real part of the loop
integrals appearing in the renormalized one-particle-irreducible
two-point Green functions $i \hat{\Gamma}_{\mu \nu}$ which can be
written as
\begin{eqnarray}
i \hat{\Gamma}_{\mu \nu}^{WW}(k) &=& -i g_{\mu \nu}(k^2 - m_W^2)
   -i \left(g_{\mu \nu}-\frac{k_{\mu}k_{\nu}}{k^2}\right)\hat{\Sigma}_T^{WW}(k^2)
   -i \frac{k_{\mu}k_{\nu}}{k^2}\hat{\Sigma}_L^{WW}(k^2) \nb \\
i \hat{\Gamma}_{\mu \nu}^{ab}(k) &=& -i g_{\mu \nu}(k^2-m_a^2)\delta_{ab}
   -i \left(g_{\mu \nu}-\frac{k_{\mu}k_{\nu}}{k^2}\right)\hat{\Sigma}_T^{ab}(k^2)
   -i \frac{k_{\mu}k_{\nu}}{k^2}\hat{\Sigma}_L^{ab}(k^2)
\end{eqnarray}
Here, $a, b = \gamma, Z$. From the above equations, we obtain \cite{denner}
\begin{eqnarray}
\delta m_W^2 &=& \tilde{Re}\Sigma_T^{WW}(m_W^2),~~~~~~~~~~~~~~~~~
\delta m_Z^2 = \tilde{Re}\Sigma_T^{ZZ}(m_Z^2) \nb \\
\delta Z_W &=& -\tilde{Re}\frac{\partial
\Sigma_T^{WW}(k^2)}{\partial k^2}\mid_
             {k^2=m_W^2} \nb \\
\delta Z_e &=& -\frac{1}{2}\delta Z_{\gamma \gamma}-\frac{s_W}{c_W}
\frac{1}{2}\delta Z_{Z \gamma}=
   \frac{1}{2}\frac{\partial \Sigma_T^{\gamma \gamma}(k^2)}{\partial k^2}\mid_
   {k^2=0}-\frac{s_W}{c_W}\frac{\Sigma_T^{\gamma \gamma}(0)}{m_Z^2},
\end{eqnarray}
in which $\Sigma$'s are the corresponding unrenormalized self energies.
\par
By using the equation $\cos^2 \theta_W = m_W^2/m_Z^2$, we find
\begin{equation}
\delta \cos \theta_W=\frac{1}{2}\cos \theta_W \left(\frac{\delta m_W^2}{m_W^2}
                                  - \frac{\delta m_Z^2}{m_Z^2}\right).
\end{equation}
\par
(b) The renormalized one-particle-irreducible two-point Green
functions of fermions(here they are charginos and neutralinos) can
be decomposed as
\begin{equation}
i \Gamma^f_{ij}(p) = i \delta_{ij}(\rlap/{p} - m_{f_i}) + i \left[ \rlap/{p}P_L
\hat{\Sigma}^L_{ij}+\rlap/{p}P_R
\hat{\Sigma}^R_{ij}+ P_L \hat{\Sigma}^{SL}_{ij}+P_R \hat{\Sigma}^{SR}_{ij}\right].
\end{equation}
where $\hat{\Sigma}^{L,R,SL,SR}_{ij}$ are the renormalized self-energy
matrices of fermions. By imposing the on-mass-shell renormalization conditions
for fermions \cite{denner}
\begin{eqnarray}
&& \lim_{p^2 \rightarrow m_{f_i}^2}\frac{1}{i (\rlap/{p} - m_{f_{i}})}
\tilde{Re}i \hat{\Gamma}_{i i}^f(p)u_{i}(p)=u_{i}(p)~~~~~~~
\tilde{Re}i \hat{\Gamma}_{i j}^f(p)u_j(p)\mid_{p^2=m_{f_j}^2}=0 \nb \\
&& \lim_{p^2 \rightarrow m_{f_i}^2}\bar{u}_{i}(p)\tilde{Re}
i \hat{\Gamma}_{i i}^f(p)\frac{1}{i (\rlap/{p} - m_{f_{i}})}=\bar{u}_{i}(p)
~~~~~~~
\tilde{Re}\bar{u}_{i}(p)i \hat{\Gamma}_{i j}^f(p)\mid_{p^2=m_{f_i}^2}=0
\end{eqnarray}
we can get the renormalization constants of fermions \cite{denner} \cite{yamada1} \cite{yamada2}:
\begin{eqnarray}
\delta m_i &=& \frac{1}{2}Re\left[m_i(\Sigma^L_{ii}(m_i^2)+\Sigma^R_{ii}(m_i^2))+
\Sigma^{SL}_{ii}(m_i^2)+ \Sigma^{SR}_{ii}(m_i^2)\right], \nb \\
\delta Z^L_{ii}&=&-Re\left[\Sigma^L_{ii}(m_i^2) + m_i^2(\Sigma^{L'}_{ii}(m_i^2)+\Sigma^{R'}_{ii}(m_i^2)) +
                         m_i (\Sigma^{SL'}_{ii}(m_i^2) + \Sigma^{SR'}_{ii}(m_i^2))\right], \nb \\
\delta Z^R_{ii}&=&-Re\left[\Sigma^R_{ii}(m_i^2) + m_i^2(\Sigma^{L'}_{ii}(m_i^2)+\Sigma^{R'}_{ii}(m_i^2)) +
                         m_i (\Sigma^{SL'}_{ii}(m_i^2) + \Sigma^{SR'}_{ii}(m_i^2))\right], \nb \\
\delta Z^L_{ij}&=&\frac{2}{m_i^2 - m_j^2}Re\left[m_j^2\Sigma^L_{ij}(m_j^2)+ m_i m_j\Sigma^{R}_{ij}(m_j^2)+
                          m_i \Sigma^{SL}_{ij}(m_j^2) + m_j \Sigma^{SR}_{ij}(m_j^2) \right], ~~~\rm{for}~ i \neq j\nb \\
\delta Z^R_{ij}&=&\frac{2}{m_i^2 - m_j^2}Re\left[m_i m_j\Sigma^L_{ij}(m_j^2)+ m_j^2\Sigma^{R}_{ij}(m_j^2)+
                          m_j \Sigma^{SL}_{ij}(m_j^2) + m_i \Sigma^{SR}_{ij}(m_j^2) \right]. ~~~\rm{for}~ i \neq j\nb \\
\end{eqnarray}
For radiative corrections to chargino and neutralino masses, we can refer
to Ref.\cite{pierce}.
\par
(c) $\delta U, \delta V$ and $\delta N$ are fixed by demanding that the wave
function corrections are symmetric \cite{yamada1} \cite{yamada2}:
\begin{eqnarray}
\delta U = \frac{1}{4}\left(\delta Z_+^R-\delta Z_+^{R \dag}\right)U \nb \\
\delta V = \frac{1}{4}\left(\delta Z_+^L-\delta Z_+^{L \dag}\right)V \nb \\
\delta N = \frac{1}{4}\left(\delta Z_0^L-\delta Z_0^{L \dag}\right)N
\end{eqnarray}
The chargino and neutralino self-energies can be found in
Ref.\cite{yamada2}.
\par
By substituting Eq.(\ref{uvn renor}) into the Lagrangian (\ref{L
for interaction}), we can obtain the counter-term of vertex
$\bar{\tilde{\chi}}_i^0 \tilde{\chi}_j^+ W^-$:
\begin{eqnarray}
\label{counterterms}
\delta {\cal L}_{\tilde{\chi}^+ \tilde{\chi}^0 W} &=&
   g W_{\mu}^- \bar{\tilde{\chi}}_i^0 \gamma^{\mu}
   \left\{
      P_L \left[
          \delta O^L_{i j}
          +O^L_{i j} \left( \frac{\delta g}{g}+\frac{1}{2} \delta Z_W \right)
      +\frac{1}{2} \sum_{k=1}^2 O^L_{i k} \delta Z^L_{+,k j}
      \right.\right. \nb \\
&&    \left.
          +\frac{1}{2} \sum_{k=1}^4 O^L_{k j} \delta Z^{L *}_{0,k i}
      \right]
     +P_R \left[
          \delta O^R_{i j}
          +O^R_{i j} \left( \frac{\delta g}{g}+\frac{1}{2} \delta Z_W \right)
      +\frac{1}{2} \sum_{k=1}^2 O^R_{i k} \delta Z^R_{+,k j}
      \right. \nb \\
&&    \left.\left.
          +\frac{1}{2} \sum_{k=1}^4 O^R_{k j} \delta Z^{R *}_{0,k i}
          \right]
   \right\}
   \tilde{\chi}_j^+ + h.c. \nb \\
\end{eqnarray}
where
\begin{eqnarray}
\delta O^L_{i j} &=& -\frac{1}{\sqrt{2}}(\delta N_{i 4} V_{j 2}^*
                                        +N_{i 4} \delta V_{j 2}^*)
                 +(\delta N_{i 2} V_{j 1}^* + N_{i 2} \delta V_{j 1}^*) \nb \\
\delta O^R_{i j} &=&  \frac{1}{\sqrt{2}}(\delta N_{i 3}^* U_{j 2}
                                        +N_{i 3}^* \delta U_{j 2})
                 +(\delta N_{i 2}^* U_{j 1} + N_{i 2}^* \delta U_{j 1})
\end{eqnarray}
For convenience, we define
\begin{eqnarray}
-i \delta \Lambda^L_{j i} &=&
          O^L_{i j} \left( \frac{\delta g}{g}+\frac{1}{2} \delta Z_W \right)
      +\frac{1}{2} \sum_{k=1}^2 O^L_{i k} \delta Z^L_{+,k j}
      +\frac{1}{2} \sum_{k=1}^4 O^L_{k j} \delta Z^{L *}_{0,k i}
      +\delta O^L_{i j} \nb \\
-i \delta \Lambda^R_{j i} &=&
          O^R_{i j} \left( \frac{\delta g}{g}+\frac{1}{2} \delta Z_W \right)
      +\frac{1}{2} \sum_{k=1}^2 O^R_{i k} \delta Z^R_{+,k j}
      +\frac{1}{2} \sum_{k=1}^4 O^R_{k j} \delta Z^{R *}_{0,k i}
      +\delta O^R_{i j}
\end{eqnarray}
Then the contribution from the counter-term to the amplitude of
the decay $\tilde{\chi}_1^+ \rightarrow W^+ \tilde{\chi}_1^0$ can
be written as
\begin{equation}
\delta{\cal M}_c = \bar{u}_{\tilde{\chi}_1^0}(k_1) \gamma^{\mu}
            (\delta \Lambda^L_{1 1} P_L + \delta \Lambda^R_{1 1} P_R)
         u_{\tilde{\chi}_1^+}(p_1) \epsilon_{\mu}(k_2)
\end{equation}
Up to the one-loop level, the total amplitude of this process is
\begin{equation}
{\cal M}_{tot} = {\cal M}_{tree} + \delta{\cal M}_v + \delta{\cal M}_c \nb
\end{equation}
where $\delta{\cal M}_v$ represents the vertex correction. For
regularization of the ultraviolet divergences in the virtual loop
corrections, we adopt the dimensional reduction scheme (DR)
\cite{DR}, which is commonly used in the calculations of the
electroweak corrections in framework of the MSSM as it preserves
supersymmetry at least at one-loop level. As we expected, the
divergence in $\delta{\cal M}_v$ will be cancelled by that in the
counter-term $\delta{\cal M}_c$ exactly. We can check this
property of renormalization both analytically and numerically.
\par
The calculations of the vertex correction $\delta{\cal M}_v$ contributed
by the one-loop diagrams in Fig.1(c) tell us that
\begin{equation}
\label{vertex correction}
\delta{\cal M}_v = \bar{u}_{\tilde{\chi}_1^0}(k_1)
        [\gamma^{\mu}(\Lambda^L_{1 1} P_L + \Lambda^R_{1 1} P_R)
    + p_1^{\mu}  (\Pi^L_{1 1} P_L + \Pi^R_{1 1} P_R)]
         u_{\tilde{\chi}_1^+}(p_1) \epsilon_{\mu}(k_2)
\end{equation}
where $\Lambda^L_{1 1}$, $\Lambda^R_{1 1}$, $\Pi^L_{1 1}$ and
$\Pi^R_{1 1}$ are the form factors whose expressions are listed in
the Appendix. To describe the magnitude of the supersymmtric
electroweak corrections to the partial width of the decay
$\tilde{\chi}_1^+ \rightarrow W^+ \tilde{\chi}_1^0$, we introduce
a quantity named relative correction $\delta$ defined as
\begin{equation}
\delta = \frac{\Gamma_{tot}(\tilde{\chi}_1^+ \rightarrow W^+
\tilde{\chi}_1^0)
             -\Gamma_{tree}(\tilde{\chi}_1^+ \rightarrow W^+ \tilde{\chi}_1^0)}
         {\Gamma_{tree}(\tilde{\chi}_1^+ \rightarrow W^+ \tilde{\chi}_1^0)}
\nb
\end{equation}
It will be used in numerical calculations in the next section.

\section{Numerical results and conclusion}
In the following we present some numerical results for the
radiative corrections to chargino decay into the lightest
neutralino and W boson. In our numerical calculations the SM input
parameters are chosen to be $m_t = 174.3$ GeV, $m_b = 4.3$ GeV,
$m_Z = 91.188$ GeV, $m_W = 80.41$ GeV and $\alpha_{EW} = 1/128$
\cite{data}. For the SUSY sector, we choose the physical
observables to be the input parameters rather than the original
parameters in the Lagrangian (e.g. the soft-SUSY-breaking
parameters $M, M^{\prime}, M^2_{\tilde{Q}}$, $\cdot \cdot \cdot$).
In this paper we use the following set of independent parameters:
\begin{equation}
\label{parameters}
( \tan \beta, m_{\tilde{\chi}_1^+}, m_{\tilde{\chi}_2^+},
m_{\tilde{\chi}_1^0}, m_{\tilde{t}_1}, \theta_t, m_{\tilde{b}_1},
m_{\tilde{b}_2}, \theta_b)
\end{equation}
The parameters $\tan \beta, m_{\tilde{\chi}_1^+},
m_{\tilde{\chi}_2^+}$ and $m_{\tilde{\chi}_1^0}$ among these nine
input parameters are used to determine the chargino and neutralino
sectors and the rest five parameters to determine the squark
sectors. The parameters $\mu$ and $M$ in the chargino matrix $X$ (
see eq.(\ref{mxn}) ) can be extracted from the input chargino
masses $m_{\tilde{\chi}_{1,2}^+}$, but their values are not
uniquely determined.
For the fixed chargino masses, we can obtain four sets of results: \\
case 1: $-\mu>M>0$, case2: $M>-\mu>0$, case 3: $M>\mu>0$ and case
4: $\mu>M>0$. The only differences among the four cases present in
their $U$ and $V$ matrices. All the four cases are considered in
this paper for completeness. The value of $m_{\tilde{t}_2}$ is
determined by $SU(2)$ gauge invariance. For all the numerical
calculations in this paper, we take $m_{\tilde{\chi}_2^+}=550$
GeV, $m_{\tilde{b}_1}=400$ GeV, $m_{\tilde{b}_2}=450$ GeV and
$m_{\tilde{t}_1}=140$ GeV.
\par
In Fig.2 we present the relative correction $\delta$ as a function
of $\tan\beta$ for $\tilde{\chi}_1^+ \rightarrow W^+
\tilde{\chi}_1^0$, assuming $m_{\tilde{\chi}_1^+}=190$ GeV,
$m_{\tilde{\chi}_1^0}=90$ GeV and $(\theta_b, \theta_t)=$(-0.3,
0.78), (-0.2, 0.4), (0.2, -0.4), (0.35, -0.4) for case 1, 2, 3 and
4 respectively. As shown in this figure, the SUSY-EW corrections
are important and especially sensitive to the value of $\tan\beta$
for case 1 and 4. For case 1 the relative correction $\delta$ is
positive when $\tan\beta \gtrsim 4$. It varies from $-5 \%$ to
$11.9 \%$ as the increment of $\tan\beta$ from 2 to 9 and
decreases slowly as $\tan\beta$ grows in the region where
$\tan\beta > 9$. The peak at $\tan\beta \sim 9$ on this curve is
due to the fact that the tree-level partial width of this decay
$\Gamma_{tree}$ has a minimum value around this position. For case
4 the relative correction is negative, large ($|\delta|>11\%$) and
increases to $-16.6 \%$ as $\tan\beta \sim 36$. For case 2 and 3,
the relative corrections are insensitive to $\tan\beta$. They are
all negative and vary from $-8.4 \%$ to $-10 \%$ and $-9.8 \%$ to
$-11.1 \%$, respectively.
\par
Fig.3(a) and (b) give the relative corrections as the functions of
$m_{\tilde{\chi}_1^+}$ in the case of $m_{\tilde{\chi}_1^0}$ = 60
GeV and 90 GeV, respectively, assuming $\tan\beta = 7$. The values
of $(\theta_b, \theta_t)$ are the same as in Fig.2. Both in the
case of $m_{\tilde{\chi}_1^0}$ = 60 GeV and 90 GeV, the relative
corrections are insensitive with $m_{\tilde{\chi}_1^+}$ for case 2
and 3. The variations of them are all less than $1 \%$.
Furthermore, the curves for case 2, 3 and 4 show that the relative
corrections in Fig.3(a) are almost the same as in Fig.3(b) for a
fixed $m_{\tilde{\chi}_1^+}$. This conclusion can be obtained from
Fig.4 more evidently. The corrections are about $-9 \%$ and $-10
\%$ for case 2 and 3 respectively and vary from $-13.7 \%$ to
about $-10 \%$ for case 4. For case 1, the relative corrections
range between $9 \%$ $(11.3 \%)$ and $3.7 \%$ $(1.7 \%)$ with
$m_{\tilde{\chi}_1^+}$ varying in the mass region 150 (180) GeV to
270 (280) GeV when $m_{\tilde{\chi}_1^0}$ = 60 (90) GeV.
\par
In Fig.4 we show graphically the dependence between the relative
correction and $m_{\tilde{\chi}_1^0}$. Here we take $\tan\beta =
7$, $m_{\tilde{\chi}_1^+}$ = 190 GeV. $\theta_b$ and $\theta_t$
are set to be the same as in figure 2. It is clear that the
relative corrections are rather stable as $m_{\tilde{\chi}_1^0}$
running from 60 GeV to 100 GeV for case 2, 3, 4 and are also not
very insensitive to $m_{\tilde{\chi}_1^0}$ for case 1. It can
reach about $11.7 \%$ and $-13.4 \%$ for case 1 and 4,
respectively.
\par
The relative corrections as the functions of the squark mixing
angles $\theta_b$ and $\theta_t$ are displayed in Fig.5 and Fig.6,
respectively. The insensitivity of the relative correction
$\delta$ to $\theta_b$ is due to the fact that the sbottoms
$\tilde{b}_{1,2}$ ($m_{\tilde{b}_{1,2}}$ = 400, 450 GeV) are much
heavier than the lighter stop $\tilde{t}_1$ ($m_{\tilde{t}_1}$ =
140 GeV). Therefore, the sbottom quark are almost decoupled and
the variation of $\theta_b$ does not affect the relative
correction too much. In contrast to the sbottom mixing angle
$\theta_b$, the stop mixing angle $\theta_t$ plays a crucial role,
as seen in Fig.6. The SUSY-EW corrections decrease or increase the
partial decay width of $\tilde{\chi}_1^+ \rightarrow W^+
\tilde{\chi}_1^0$ significantly depending on $\theta_t$.
\par
In summary, we have computed the supersymmetric electroweak
corrections to the partial width of the deacy $\tilde{\chi}_1^+
\rightarrow W^+ \tilde{\chi}_1^0$ in the MSSM. Only the third
generation of quark and squark are considered because of the
Yukawa-like couplings of the relevant vertices. These corrections
are not very sensitive to the mass of the lightest neutralino and
the sbottom mixing angle, but depend strongly on the parameters
$\tan\beta$ and $\theta_t$. The magnitude of the corrections can
exceed $10 \%$ in some parameter space, therefore they should be
taken into account in any reliable analysis.

\noindent{\large\bf Acknowledgements:} This work was supported in
part by the National Natural Science Foundation of China(project
Nos: 19875049, 10005009), the Doctoral Program Foundation of the
Education Ministry of China, and a grant from the Ministry of
Science and Technology of China.

\section{Appendix}
In this appendix, we list the form factors of the decay
$\tilde{\chi}_i^+ \rightarrow W^+ \tilde{\chi}_j^0$ (i=1,2,
j=1,2,3,4). The form factors $\Lambda^{L,R}_{i j}$ and
$\Pi^{L,R}_{i j}$ can be divided into four parts respectively
\begin{eqnarray}
\Lambda^{L,R}_{i j} &=& \Lambda^{(1)L,R}_{i j} + \Lambda^{(2)L,R}_{i j}
                      + \Lambda^{(3)L,R}_{i j} + \Lambda^{(4)L,R}_{i j} \nb \\
\Pi^{L,R}_{i j} &=& \Pi^{(1)L,R}_{i j} + \Pi^{(2)L,R}_{i j}
                  + \Pi^{(3)L,R}_{i j} + \Pi^{(4)L,R}_{i j}
\end{eqnarray}
where $\Lambda^{(a)L,R}_{i j}$ and $\Pi^{(a)L,R}_{i j}$
(a=1,2,3,4) are the form factors contributed by the four one-loop
diagrams in Fig.1($c-1 \sim c-4$), respectively. In the
calculation of these form factors, the following vertices (or
their conjugate vertices) will be used:
\begin{eqnarray}
\bar{t}-\tilde{\chi}_i^+-\tilde{b}_j &:& ~~~~~~~~~~
  V_{t \tilde{\chi}_i^+ \tilde{b}_j}^L P_L
+ V_{t \tilde{\chi}_i^+ \tilde{b}_j}^R P_R \nb \\
\bar{b}-\bar{\tilde{\chi}}_i^+-\tilde{t}_j &:& ~~~~~~~~~
 (V_{b \tilde{\chi}_i^+ \tilde{t}_j}^L P_L
+ V_{b \tilde{\chi}_i^+ \tilde{t}_j}^R P_R) C \nb \\
\bar{t}-\tilde{\chi}_i^0-\tilde{t}_j &:& ~~~~~~~~~~
  V_{t \tilde{\chi}_i^0 \tilde{t}_j}^L P_L
+ V_{t \tilde{\chi}_i^0 \tilde{t}_j}^R P_R \nb \\
\bar{b}-\tilde{\chi}_i^0-\tilde{b}_j &:& ~~~~~~~~~~
  V_{b \tilde{\chi}_i^0 \tilde{b}_j}^L P_L
+ V_{b \tilde{\chi}_i^0 \tilde{b}_j}^R P_R \nb \\
\tilde{b}_i^*-\tilde{t}_j-W^- &:& ~~~~~~~~~~
  V_{\tilde{b}_i \tilde{t}_j W} (p + p^{\prime})^{\mu}
\end{eqnarray}
where $p$ and $p^{\prime}$ are the incoming momentum of $\tilde{t}_j$
and outgoing momentum of $\tilde{b}_i$ respectively.
The explicit expressions of these vertices can be found in Ref.\cite{mssm2}.
and \cite{haber}.
\par
Now we can write down the form factors of the decay
$\tilde{\chi}_i^+(p_1) \rightarrow \tilde{\chi}_j^0(k_1) + W^+(k_2)$ :
\begin{eqnarray}
\label{forfac}
\Lambda^{(1)L}_{i j} &=& \frac{3}{8 \pi^2} \sum_{l, k=1}^{2}
    V_{b \tilde{\chi}_i^+ \tilde{t}_l}^{R*}
    V_{b \tilde{\chi}_j^0 \tilde{b}_k}^{R}
    V_{\tilde{b}_k \tilde{t}_l W}
    C_{24}^{(1)} \nb \\
\Pi^{(1)L}_{i j} &=& -\frac{3}{8 \pi^2} \sum_{l, k=1}^{2}
    V_{\tilde{b}_k \tilde{t}_l W}
        [(C_{11}^{(1)} -C_{12}^{(1)} + C_{21}^{(1)} - C_{23}^{(1)})
         V_{b \tilde{\chi}_i^+ \tilde{t}_l}^{R*}
     V_{b \tilde{\chi}_j^0 \tilde{b}_k}^{R}
     m_{\tilde{\chi}_j^0} \nb \\
&&      +(C_{11}^{(1)} - C_{12}^{(1)})
         V_{b \tilde{\chi}_i^+ \tilde{t}_l}^{R*}
     V_{b \tilde{\chi}_j^0 \tilde{b}_k}^{L}
     m_b
        +(C_{22}^{(1)} - C_{23}^{(1)})
     V_{b \tilde{\chi}_i^+ \tilde{t}_l}^{L*}
     V_{b \tilde{\chi}_j^0 \tilde{b}_k}^{L}
     m_{\tilde{\chi}_i^+}] \nb \\
\Lambda^{(1)R}_{i j} &=& \Lambda^{(1)L}_{i j}( L \leftrightarrow R ) \nb \\
\Pi^{(1)R}_{i j} &=& \Pi^{(1)L}_{i j}( L \leftrightarrow R ) \nb \\
\Lambda^{(2)L}_{i j} &=& -\frac{3}{8 \pi^2} \sum_{l, k=1}^{2}
    V_{t \tilde{\chi}_i^+ \tilde{b}_k}^{L}
    V_{t \tilde{\chi}_j^0 \tilde{t}_l}^{L*}
    V_{\tilde{b}_k \tilde{t}_l W} C_{24}^{(2)} \nb \\
\Pi^{(2)L}_{i j} &=& \frac{3}{8 \pi^2} \sum_{l, k=1}^{2}
    V_{\tilde{b}_k \tilde{t}_l W}
        [(C_{11}^{(2)} - C_{12}^{(2)} + C_{21}^{(2)} - C_{23}^{(2)})
     V_{t \tilde{\chi}_i^+ \tilde{b}_k}^{L}
     V_{t \tilde{\chi}_j^0 \tilde{t}_l}^{L*}
     m_{\tilde{\chi}_j^0} \nb \\
&&      +(C_{11}^{(2)} - C_{12}^{(2)})
         V_{t \tilde{\chi}_i^+ \tilde{b}_k}^{L}
     V_{t \tilde{\chi}_j^0 \tilde{t}_l}^{R*}
     m_t
    +(C_{22}^{(2)} - C_{23}^{(2)})
     V_{t \tilde{\chi}_i^+ \tilde{b}_k}^{R}
     V_{t \tilde{\chi}_j^0 \tilde{t}_l}^{R*}
     m_{\tilde{\chi}_i^+}] \nb \\
\Lambda^{(2)R}_{i j} &=& \Lambda^{(2)L}_{i j}( L \leftrightarrow R ) \nb \\
\Pi^{(2)R}_{i j} &=& \Pi^{(2)L}_{i j}( L \leftrightarrow R ) \nb \\
\Lambda^{(3)L}_{i j} &=& -i \frac{3 g}{16 \sqrt{2} \pi^2} \sum_{k=1}^{2}
    \{V_{t \tilde{\chi}_i^+ \tilde{b}_k}^{L} m_t
        [C_0^{(3)} m_b
     V_{b \tilde{\chi}_j^0 \tilde{b}_k}^{L*}
        -C_{11}^{(3)} m_{\tilde{\chi}_j^0}
     V_{b \tilde{\chi}_j^0 \tilde{b}_k}^{R*}] \nb \\
&&   +V_{t \tilde{\chi}_i^+ \tilde{b}_k}^{R} m_{\tilde{\chi}_i^+}
        [(C_0^{(3)} + C_{12}^{(3)}) m_b
     V_{b \tilde{\chi}_j^0 \tilde{b}_k}^{L*}
        -(C_{11}^{(3)} - C_{12}^{(3)}) m_{\tilde{\chi}_j^0}
     V_{b \tilde{\chi}_j^0 \tilde{b}_k}^{R*}] \} \nb \\
\Lambda^{(3)R}_{i j} &=& i \frac{3 g}{16 \sqrt{2} \pi^2} \sum_{k=1}^{2}
    \{C_{12}^{(3)}
      m_t m_{\tilde{\chi}_i^+}
      V_{t \tilde{\chi}_i^+ \tilde{b}_k}^{L}
      V_{b \tilde{\chi}_j^0 \tilde{b}_k}^{R*}
    -(C_0^{(3)} + C_{11}^{(3)})
      m_b m_{\tilde{\chi}_j^0}
      V_{t \tilde{\chi}_i^+ \tilde{b}_k}^{R}
      V_{b \tilde{\chi}_j^0 \tilde{b}_k}^{L*} \nb \\
&&        +[\frac{1}{2} - 2 C_{24}^{(3)}
          +(C_{12}^{(3)} + C_{23}^{(3)}) m_W^2
      +(C_{22}^{(3)} - C_{23}^{(3)}) m_{\tilde{\chi}_i^+}^2 \nb \\
&&    +(C_{11}^{(3)} - C_{12}^{(3)} + C_{21}^{(3)} -C_{23}^{(3)})
        m_{\tilde{\chi}_j^0}^2]
      V_{t \tilde{\chi}_i^+ \tilde{b}_k}^{R}
      V_{b \tilde{\chi}_j^0 \tilde{b}_k}^{R*} \} \nb \\
\Pi^{(3)L}_{i j} &=& -i \frac{3 g}{8 \sqrt{2} \pi^2} \sum_{k=1}^{2}
    [ C_{12}^{(3)} m_t
      V_{t \tilde{\chi}_i^+ \tilde{b}_k}^{L}
      +(C_{22}^{(3)} - C_{23}^{(3)}) m_{\tilde{\chi}_i^+}
      V_{t \tilde{\chi}_i^+ \tilde{b}_k}^{R}]
      V_{b \tilde{\chi}_j^0 \tilde{b}_k}^{R*} \nb \\
\Pi^{(3)R}_{i j} &=& i \frac{3 g}{8 \sqrt{2} \pi^2} \sum_{k=1}^{2}
    V_{t \tilde{\chi}_i^+ \tilde{b}_k}^{R}
    [ (C_0^{(3)} + C_{11}^{(3)}) m_b
      V_{b \tilde{\chi}_j^0 \tilde{b}_k}^{L*}
      -(C_{11}^{(3)} - C_{12}^{(3)} + C_{21}^{(3)} - C_{23}^{(3)})
      m_{\tilde{\chi}_j^0}
      V_{b \tilde{\chi}_j^0 \tilde{b}_k}^{R*}
    ] \nb \\
\Lambda^{(4)L}_{i j} &=& -i \frac{3 g}{16 \sqrt{2} \pi^2} \sum_{k=1}^{2}
    \{C_{12}^{(4)}
     m_b m_{\tilde{\chi}_i^+}
     V_{b \tilde{\chi}_i^+ \tilde{t}_k}^{L*}
     V_{t \tilde{\chi}_j^0 \tilde{t}_k}^{R}
    -(C_0^{(4)} + C_{11}^{(4)})
     m_t m_{\tilde{\chi}_j^0}
     V_{b \tilde{\chi}_i^+ \tilde{t}_k}^{R*}
     V_{t \tilde{\chi}_j^0 \tilde{t}_k}^{L} \nb \\
&&        +[ \frac{1}{2} - 2 C_{24}^{(4)}
      +(C_{12}^{(4)} + C_{23}^{(4)}) m_W^2
      +(C_{22}^{(4)} - C_{23}^{(4)}) m_{\tilde{\chi}_i^+}^2 \nb \\
&&        +(C_{11}^{(4)} - C_{12}^{(4)} + C_{21}^{(4)} - C_{23}^{(4)})
            m_{\tilde{\chi}_j^0}^2]
     V_{b \tilde{\chi}_i^+ \tilde{t}_k}^{R*}
     V_{t \tilde{\chi}_j^0 \tilde{t}_k}^{R} \} \nb \\
\Lambda^{(4)R}_{i j} &=& i \frac{3 g}{16 \sqrt{2} \pi^2} \sum_{k=1}^{2}
    \{V_{b \tilde{\chi}_i^+ \tilde{t}_k}^{L*} m_b
          [C_0^{(4)} m_t
       V_{t \tilde{\chi}_j^0 \tilde{t}_k}^{L}
      -C_{11}^{(4)} m_{\tilde{\chi}_j^0}
       V_{t \tilde{\chi}_j^0 \tilde{t}_k}^{R}] \nb \\
&&   +V_{b \tilde{\chi}_i^+ \tilde{t}_k}^{R*} m_{\tilde{\chi}_i^+}
          [(C_0^{(4)} + C_{12}^{(4)}) m_t
       V_{t \tilde{\chi}_j^0 \tilde{t}_k}^{L}
      -(C_{11}^{(4)} - C_{12}^{(4)}) m_{\tilde{\chi}_j^0}
       V_{t \tilde{\chi}_j^0 \tilde{t}_k}^{R}] \} \nb \\
\Pi^{(4)L}_{i j} &=& -i \frac{3 g}{8 \sqrt{2} \pi^2} \sum_{k=1}^{2}
     V_{b \tilde{\chi}_i^+ \tilde{t}_k}^{R*}
          [(C_0^{(4)} + C_{11}^{(4)}) m_t
       V_{t \tilde{\chi}_j^0 \tilde{t}_k}^{L}
      -(C_{11}^{(4)} - C_{12}^{(4)} + C_{21}^{(4)} - C_{23}^{(4)})
       m_{\tilde{\chi}_j^0}
       V_{t \tilde{\chi}_j^0 \tilde{t}_k}^{R}] \nb \\
\Pi^{(4)R}_{i j} &=& i \frac{3 g}{8 \sqrt{2} \pi^2} \sum_{k=1}^{2}
      [C_{12}^{(4)} m_b
       V_{b \tilde{\chi}_i^+ \tilde{t}_k}^{L*}
      +(C_{22}^{(4)} - C_{23}^{(4)}) m_{\tilde{\chi}_i^+}
       V_{b \tilde{\chi}_i^+ \tilde{t}_k}^{R*}]
       V_{t \tilde{\chi}_j^0 \tilde{t}_k}^{R}
\end{eqnarray}
where
\begin{eqnarray}
C_{11, 12, 21, 22, 23, 24}^{(1)} &=& C_{11, 12, 21, 22, 23, 24}(k_1,
      -p_1, m_{\tilde{b}_k}, m_b, m_{\tilde{t}_l}) \nb \\
C_{11, 12, 21, 22, 23, 24}^{(2)} &=& C_{11, 12, 21, 22, 23, 24}(k_1,
      -p_1, m_{\tilde{t}_l}, m_t, m_{\tilde{b}_k}) \nb \\
C_{0, 11, 12, 21, 22, 23, 24}^{(3)} &=& C_{0, 11, 12, 21, 22, 23, 24}(k_1,
      -p_1, m_b, m_{\tilde{b}_k}, m_t) \nb \\
C_{0, 11, 12, 21, 22, 23, 24}^{(4)} &=& C_{0, 11, 12, 21, 22, 23, 24}(k_1,
      -p_1, m_t, m_{\tilde{t}_k}, m_b)
\end{eqnarray}
The definitions and numerical calculation formula of the
one-point, two-point and three-point Passarino-Veltman integrals
are adopted from Ref.\cite{cfun} and Ref.\cite{velt},
respectively. 

\vskip 10mm
\begin{flushleft} {\bf Figure Captions} \end{flushleft}

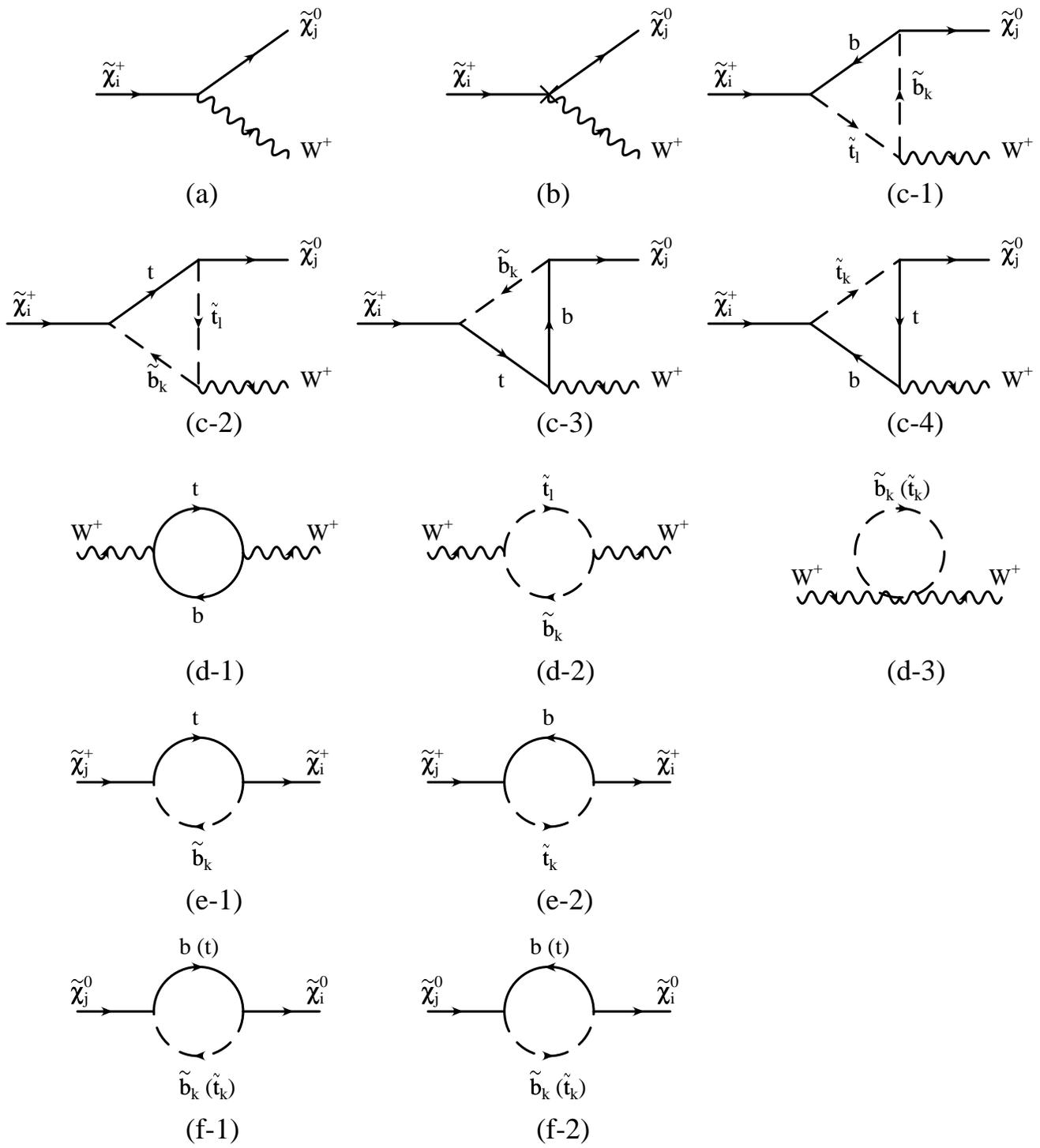
{\bf Fig.1} The relevant Feynman diagrams to the decay
$\tilde{\chi}_1^+ \rightarrow W^+ \tilde{\chi}_1^0$: (a)
tree-level diagram; (b) counter-term for vertex; (c-1)-(c-4) one-loop
vertex corrections; (d), (e) and (f) self-energies of W boson,
chargino and neutralino respectively. In diagrams (a), (b) and
(c), $i$ = $j$ = 1. In diagrams (c), (d), (e) and (f) the
subscript $k$ and $l$ can take from 1 to 2.
\par
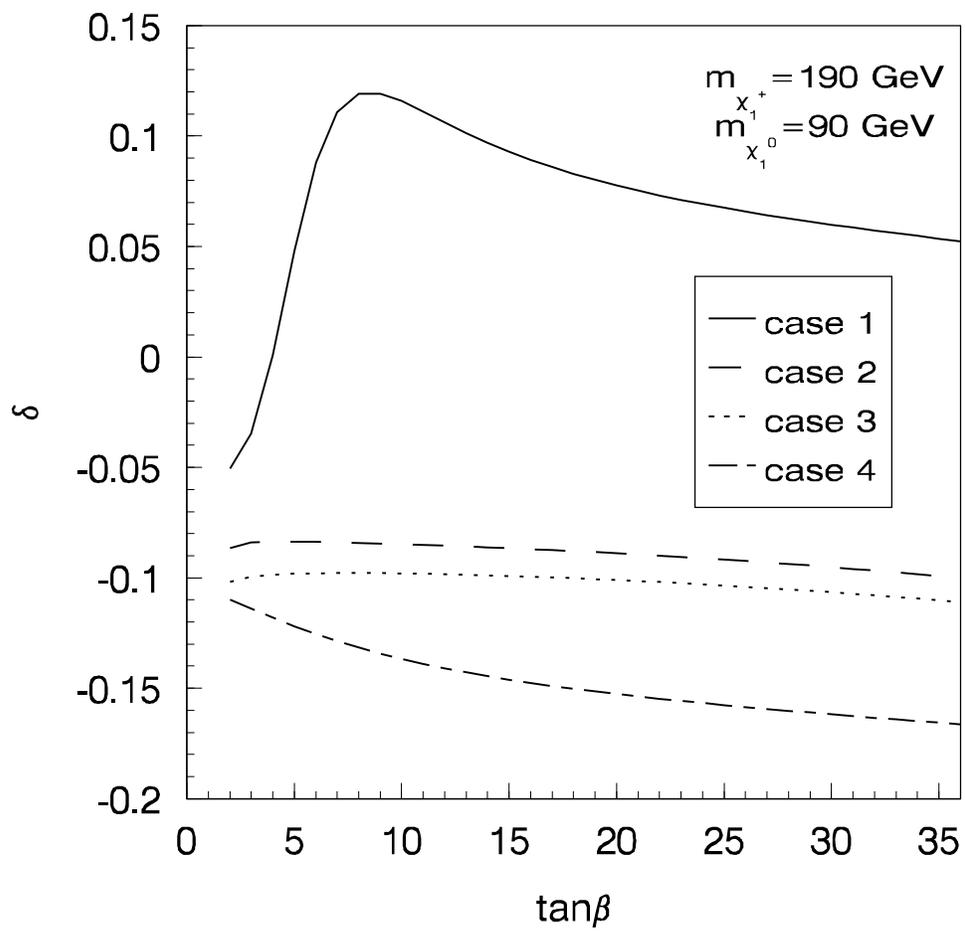
{\bf Fig.2} The relative correction as a function  of $\tan\beta$.
\par
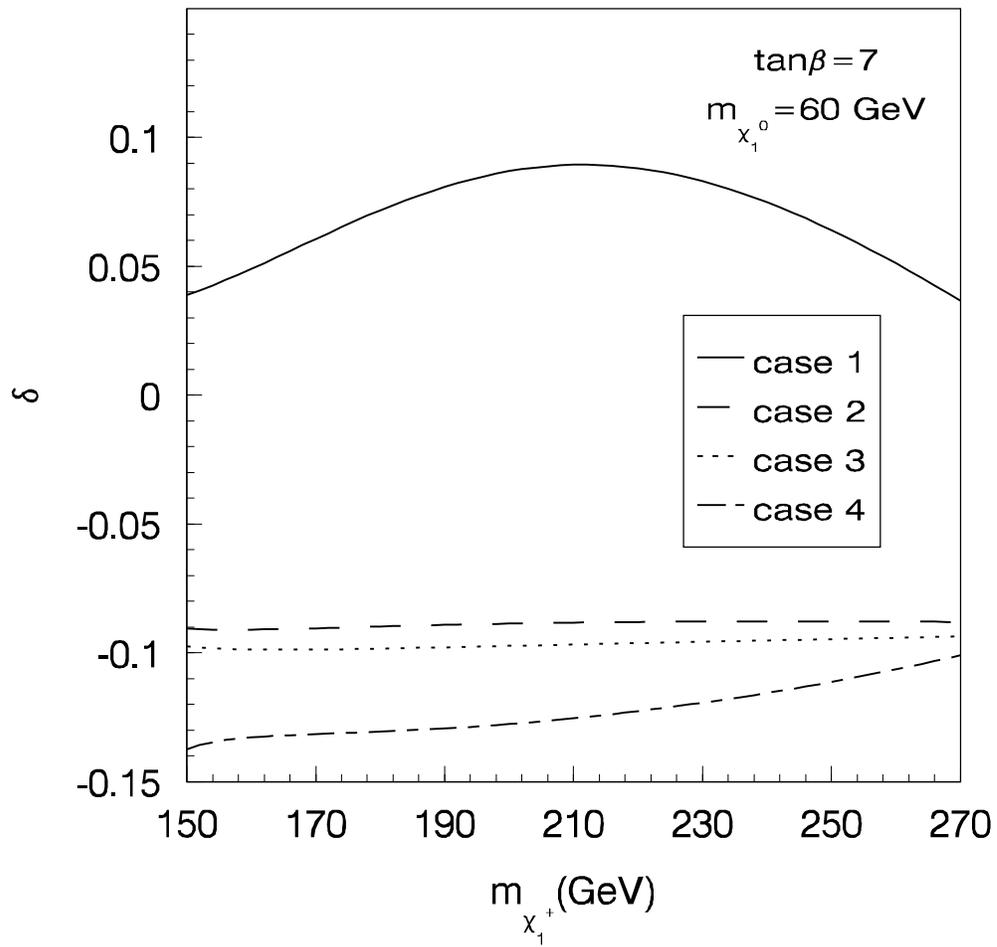
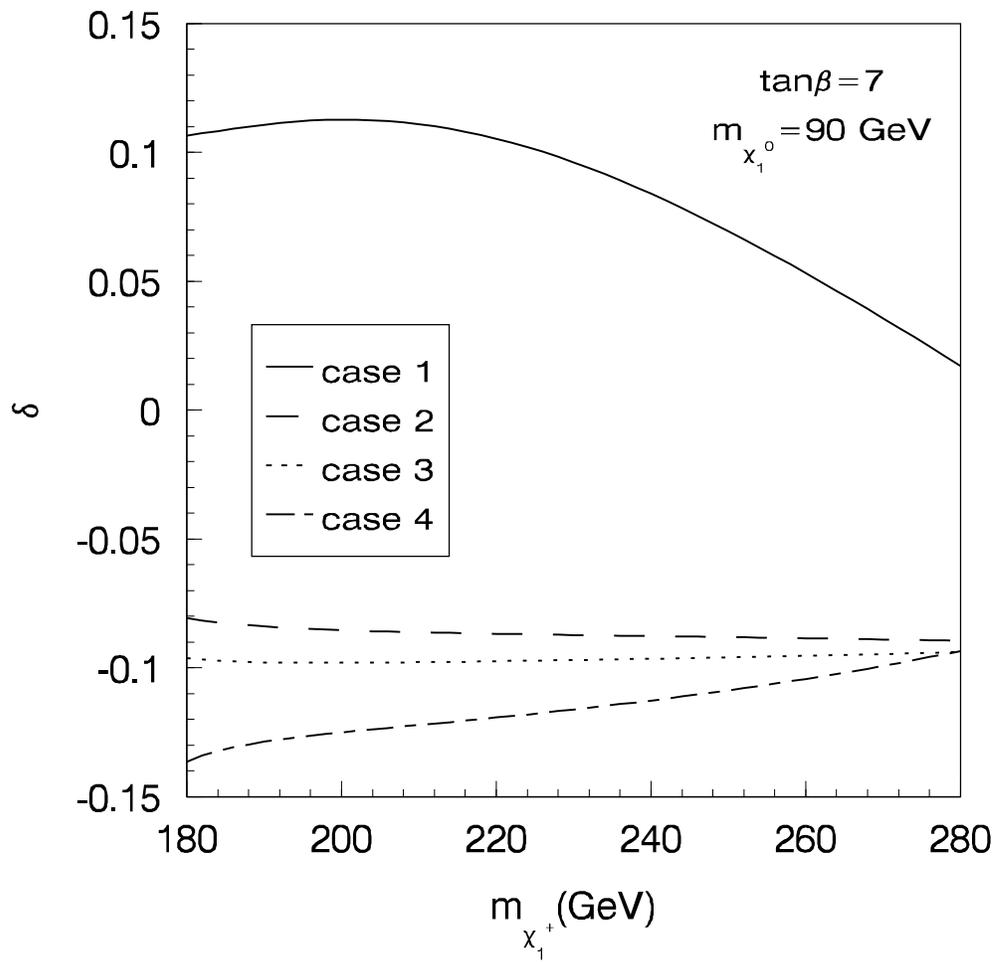
{\bf Fig.3} The relative correction as a function of $m_{\tilde{\chi}_1^+}$.
\par
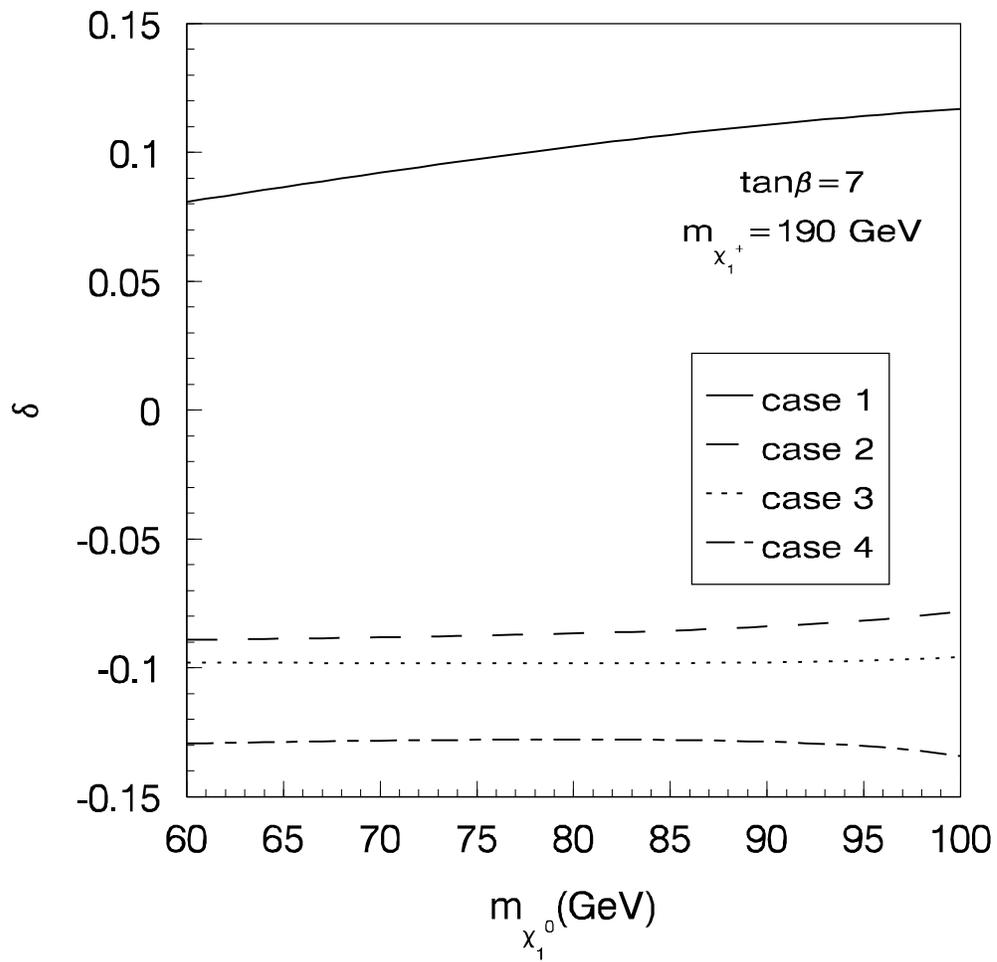
{\bf Fig.4} The relative correction as a function of $m_{\tilde{\chi}_1^0}$.
\par
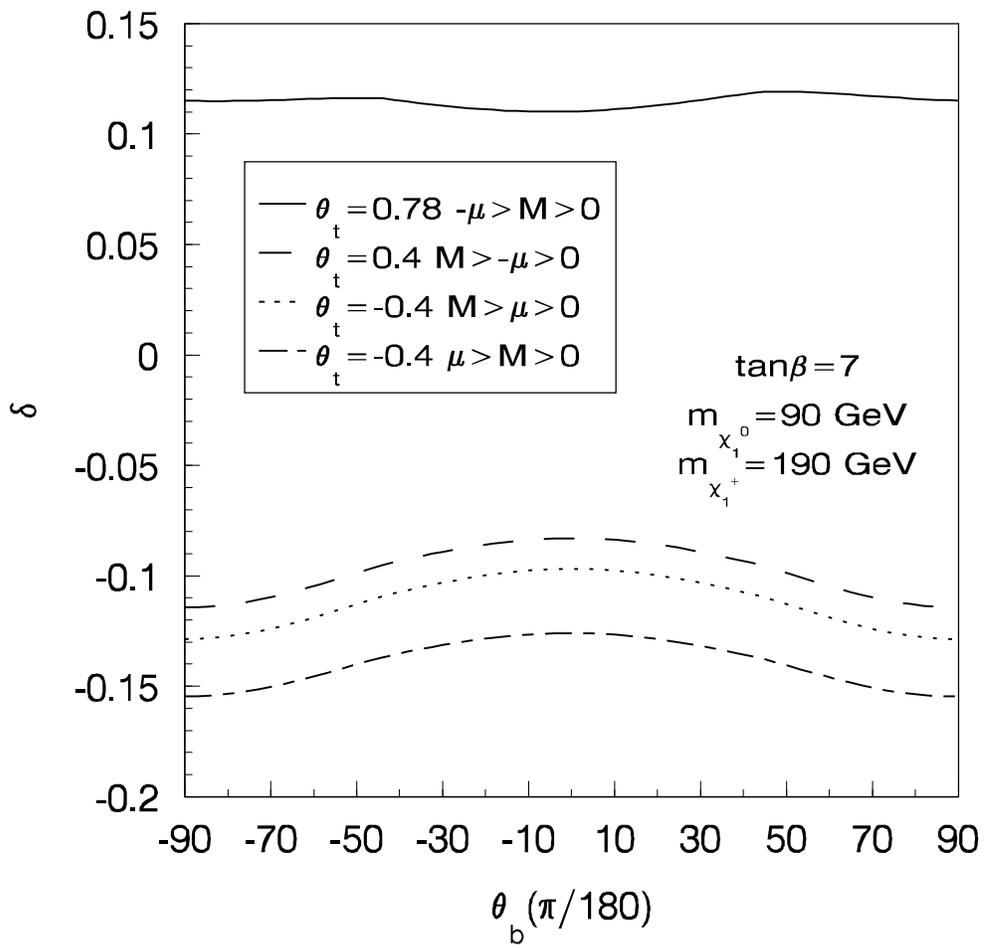
{\bf Fig.5} The relative correction as a function of $\theta_b$.
\par
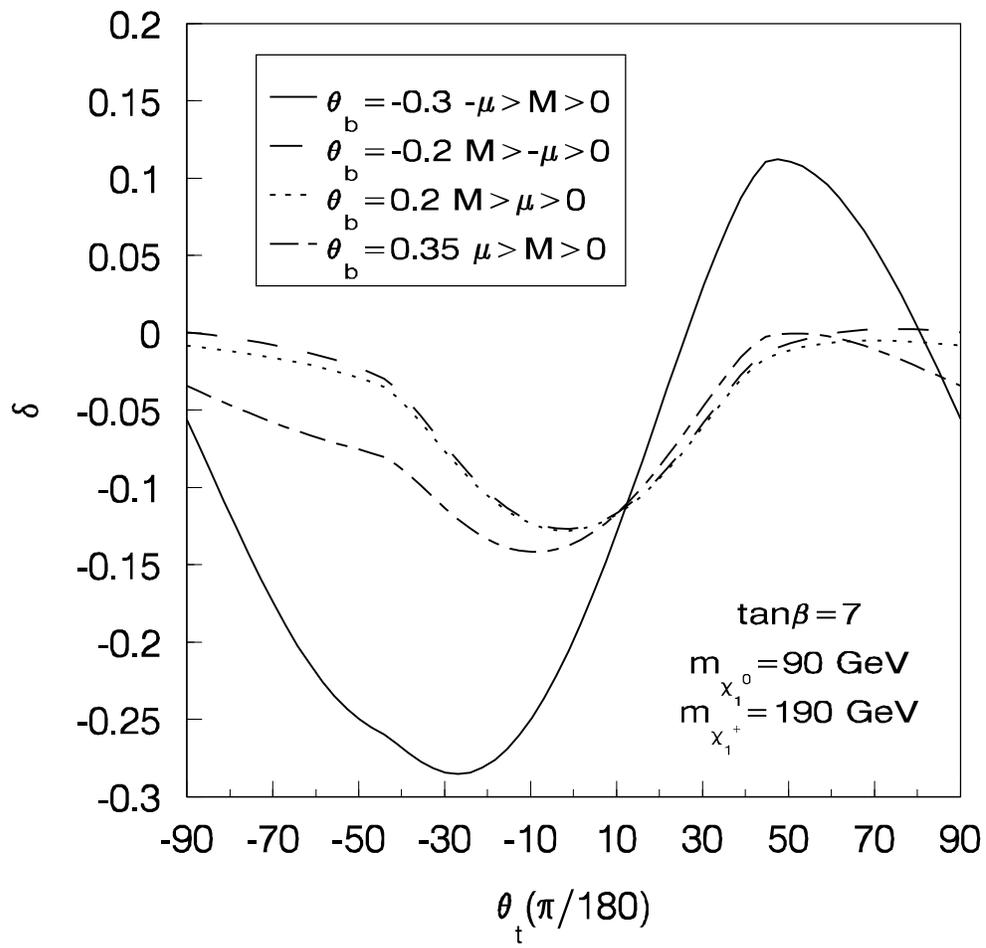
{\bf Fig.6} The relative correction as a function of $\theta_t$.
\end{document}